\documentclass[10pt,a4paper]{article}

\usepackage[T1]{fontenc}
\usepackage{comment}
\usepackage{placeins}
\usepackage[utf8]{inputenc}
\usepackage{amssymb,amsmath,amsthm,amsfonts,mathrsfs}    
\usepackage{bm}                                 
\usepackage{mathtools}                          
\usepackage{stmaryrd}                           
\usepackage{breqn}                              
\usepackage[textfont=footnotesize,labelfont={footnotesize,bf}]{caption}
\usepackage{subcaption}
\usepackage{booktabs}
\usepackage[inline]{enumitem}
\usepackage{graphicx}
\usepackage[dvipsnames]{xcolor}
\usepackage[top=30mm,right=30mm,left=30mm,bottom=30mm]{geometry}
\usepackage[colorlinks=true,linkcolor=blue,urlcolor=blue]{hyperref}
\usepackage[affil-it,auth-sc]{authblk}
\usepackage[colorinlistoftodos,textwidth=25mm,textsize=footnotesize]{todonotes}
\usepackage{lipsum}                             
\usepackage[
	backend=biber,
	style=numeric,
	citestyle=numeric-comp,
	maxbibnames=99,
	minbibnames=99,
	maxcitenames=2,
	mincitenames=1,
	giveninits=true,
	sorting=none,
	sortlocale=auto,
	natbib=true,
	url=false,
	isbn=false,
	doi=true,
	eprint=true
]{biblatex}
\graphicspath{{./}{./figures/}}                 

\DeclareMathOperator{\tr}{tr}
\DeclareMathOperator{\Rp}{Re}
\DeclareMathOperator{\Ip}{Im}

\title{ 
Effects of prestress in the coating of an elastic disk
}

\author[1]{M. Gaibotti}
\author[2]{S.G. Mogilevskaya}
\author[1]{A. Piccolroaz}
\author[1]{D. Bigoni\footnote{Corresponding author: e-mail: \href{mailto:bigoni@ing.unitn.it}{bigoni@ing.unitn.it}; phone: +39\,0461\,282507.}}

\affil[1]{Department of Civil, Environmental, and Mechanical Engineering, University of Trento, Trento, Italy}
\affil[2]{Department of Civil, Environmental and Geo-Engineering, University of Minnesota, 500 Pillsbury Drive S.E. Minneapolis, MN 55455-0116, USA}

\date{Dedicated to Professor Yibin Fu on the occasion of his 60th birthday} 

\begin{document}
\maketitle

\begin{abstract}
\noindent
An elastic disk is coated with an elastic rod, uniformly prestressed with a tensile or compressive axial force. The prestress state is assumed to be induced by three different models of external radial load or by \lq shrink-fit' forcing the coating onto the disk. 
The prestressed coating/disk system, when loaded with an additional and arbitrary {\it incremental} external load, experiences {\it incremental} displacement, strain, and stress, which are solved via complex potentials. 
The analysis incorporates models for both perfect and imperfect bonding at the coating/disk interface. 
The derived solution highlights the significant influence not only of the prestress but also of the method employed to generate it. 
These two factors lead, in different ways, to a loss or an increase in incremental stiffness for compressive or tensile prestress. The first bifurcation load of the structure (which differs for different prestress generations) is determined in a perturbative way. The results emphasize the importance of modelling the load and may find applications in flexible electronics and robot arms subject to pressure or uniformly-distributed radial forces. 
\end{abstract}



\section{Introduction}
\label{sec:introduction}

Thermo-mechanical and chemical treatments of materials such as surface hardening, welding, and contact with gears, rolling bars, or manufacturing tools often induce residual stresses in materials and components. Depending on their nature and how they interact with applied loads, these stresses can sometimes be beneficial or detrimental to stiffness and strength \cite{jensen1990decohesion,noyan1991residual,beuth1992cracking,jorgensen1995cracking,chen2022review}. In particular, internal stresses in coatings and thin films are recognized as the primary cause for loss of mechanical and adhesive properties, possibly leading to failure. In other circumstances, residual stresses can improve the mechanical performance of a piece, as happens for instance in the well-known case of tempered glass. 
As related to growth and sometimes remodelling, residual stresses are present in many biological tissues \cite{humphrey2003continuum}, and particularly in arteries \cite{holzapfel2010constitutive}, where they strongly influence the mechanical response. 

Research on soft materials for applications in tissue mechanics and soft devices has fostered research on residual stresses (defined as stress states persisting even in the absence of external loads) in nonlinear elasticity. This field has been theoretically developed in \cite{hoger1985residual, hoger1986determination}, while wave propagation (concerning the possibility of detecting residual stresses) has been analysed in \cite{armenakas1963vibrations, ogden84, man1987towards, gei, shams2011initial}. Finally, simple shear, azimuthal shear, and torsion of a cylinder have been investigated in \cite{merodio2013influence}. 

The present article investigates the response to {\it incremental} external load of a mechanical system involving a prestressed and curved element. 
The prestress can be residual stress induced by thermal loading or by a forced insertion of a piece into another, or can be generated by an external load, here considered of three different types (hydrostatic pressure, centrally directed, and dead). 
Although they generate the same prestress in the curved element, the three different loadings produce different incremental effects, thus leading to different incremental stress states when perturbed through an additional incremental load, externally applied to the system. 
The mechanical system considered here is a linear elastic disk coated by a circular elastic rod, assumed axially inextensible and prestressed in tension or compression, in an arrangement similar to that analysed in \cite{gaibotti2024bifurcations}, 
so that the treatment includes perfect and imperfect bonding (the latter meaning unprescribed slip in the tangential direction) between rod and disk. The elastic rod is modelled within the (exact) second-order theory of curved beams. The inner disk is solved via Kolosov-Muskhelishvili complex potentials, thus leading to a general solution, holding for every possible external load increment. The latter is assumed to be superimposed to a possible pre-existing radial loading, which generates the prestress state in the coating.  
The analysis shows that the prestress has a strong effect on the stiffness of the coating/disk system, which increases (decreases) for tensile (compressive) axial internal force. 
The decrease of stiffness at the increase of compressive axial force in the annular rod leads to a perturbative determination of the buckling condition obtained in \cite{gaibotti2024bifurcations} via bifurcation analysis. 
These findings are not surprising, but have previously been analytically investigated only on  geometries simpler than that considered here, typically, representing a stack of layers \cite{bigoni2008dynamics,cai2000exact,gei2002vibration}, 
or for circular geometry, but perfect bonding between disk and coating and only pressure loading \cite{ogdencino}. 
The set-up of the mechanical problem proposed here is sufficiently simple to make an analytical solution viable, which would be otherwise awkward. However, the circular geometry analysed here represents a model problem for the determination of the influence of prestress or residual stress on the behaviour of a curved elastic system. The presented results may find applications in the mechanics of coated fibres or stented arteries.

\section{The coating and the disk}
\label{sec:kinematics}

The equations governing the behaviour of the annular rod modelling the coating and of the inner elastic disk are summarized below, the interested reader can find a detailed derivation in \cite{gaibotti2024bifurcations,gaibotti2022isoDisk}.

\subsection{Statics and kinematics of a circular rod}
\label{subsec:circStatics}

The circular rod of radius $R$ considered here is modelled as axially inextensible and unshearable, involving a linear relationship between moment and curvature, ruled by the bending stiffness $B$ (equal to the product between Young's modulus, $E^{\text{c}}$, of the rod and the second moment of inertia of its cross-section, $J$). 
The elastic disk is made up of a linear isotropic elastic material deformed in plane strain or plane stress, characterized by the Kolosov constant 
\begin{equation}
\label{eq:kolosovCnst}
    \kappa^{\text{d}} = 
    3 - 4 \nu^{\text{d}} \quad \text{for plane strain,} 
    \qquad
    \kappa^{\text{d}} =
    \frac{3 - \nu^{\text{d}}}{1 + \nu^{\text{d}}} \quad \text{for plane stress,}
\end{equation}
where the superscript \lq d' stands for \lq disk', having Poisson's ratio equal to $\nu^{\text{d}}$.

\begin{figure}[htb!]
    \centering
    \includegraphics[scale=0.93,keepaspectratio]{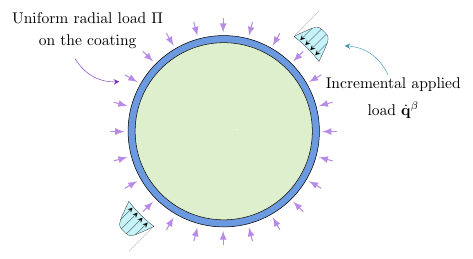}
    \includegraphics[scale=0.93,keepaspectratio]{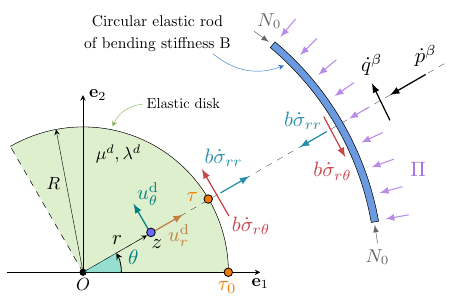}
    \caption{Left: The elastic disk coated with a circular elastic rod and subject to an external (uniform and radial) load, $\Pi$, to which an incremental load, $\dot{\mathbf{q}}^\beta$, is superimposed. Right: stress components transmitted between coating and disk. The incremental load has tangential and radial components $\dot{q}^\beta$ and $\dot{p}^\beta$ load, respectively, see equation \eqref{eq:qdotSplit}$_1$. The radial external load $\Pi$ can assume three different forms: (i.)  hydrostatic pressure, (ii.) centrally directed, and (iii.) dead. All three loads $\Pi$ may generate the same axial prestress $N_0$ in the circular rod, the difference between them only appears in the increment.}
    \label{fig:Pload}
\end{figure}

In the plane spanned by the two orthogonal unit vectors $\mathbf{e}_1$ and $\mathbf{e}_2$, the elastic rod, circular in its undeformed configuration, is assumed for the moment to undergo a large deformation. The obtained nonlinear behaviour will be reduced later to the linearized incremental response, needed to account for the presence of prestress.
The rod is  parametrized by the arc length $s$, singling out the unit tangent vector $\mathbf{t}_0$, the principal unit normal $\mathbf{n}_0$, and the curvature $\kappa_0$ at every point $\mathbf{x}_0$ of the reference configuration, together with the unit vector $\mathbf{m}_0 = \mathbf{t}_0 \times \mathbf{e}_3$, where $\mathbf{e}_3 = \mathbf{e}_1 \times \mathbf{e}_2$ is the out-of-plane unit vector. In polar coordinates, the displacement $\mathbf{u}$ and its derivative with respect to $s$ can be defined as
\begin{equation}
\label{eq:dispCirc}
    \mathbf{u} = u_{r}\,\mathbf{m}_0 + u_{\theta}\,\mathbf{t}_0 ,
    \quad
    \frac{\partial \mathbf{u}}{\partial s} = \left(\frac{\partial u_{\theta}}{\partial s} + \frac{u_r}{R}\right) \mathbf{t}_0 + \left(\frac{\partial u_r}{\partial s} - \frac{u_{\theta}}{R}\right) \mathbf{m}_0 ,
\end{equation}
so that, in the deformed configuration of the rod, the above-defined kinematic descriptors become
\begin{equation}
\label{eq:t0m0Circ}
\begin{aligned}
    &\mathbf{t} = \mathbf{t}_0 + \left(\frac{\partial u_r}{\partial s} - \frac{u_{\theta}}{R}\right) \mathbf{m}_0 , 
    &&\kappa\,\mathbf{n} = \left(\frac{1}{R}\frac{\partial u_r}{\partial s} - \frac{u_{\theta}}{R}\right) \mathbf{t}_0 - \frac{1}{R} \mathbf{m}_0, \\[3 mm]
    &\mathbf{m} = \left(\frac{u_{\theta}}{R} - \frac{\partial u_r}{\partial s}\right) \mathbf{t}_0 + \mathbf{m}_0 , 
    &&\frac{\partial \mathbf{m}}{\partial s} = \mathbf{t}_0 + \left(\frac{u_{\theta}}{R} - \frac{\partial u_r}{\partial s}\right) \mathbf{m}_0 ,
\end{aligned}
\end{equation}
where $ds = R\, d\theta$, being $\theta$ the circumferential angle measured positively in a counter-clockwise direction, as depicted in Fig.~\ref{fig:Pload}.

\subsubsection{Incremental equilibrium of the coating for three different radial loads}
\label{subsec:circDisk}

As illustrated in Fig.~\ref{fig:Pload}, the coating is modelled as a circular rod, which is subjected in its undeformed configuration to a uniform load $\Pi$, acting radially and so producing only a uniform axial prestress, $N_0$, without shear force $T_0$ and bending moment $M_0$, 
\begin{equation}
\label{eq:presB0}
    N_0 = -\Pi\, R, \quad T_{0} = M_{0} = 0 ,
\end{equation}
where $\Pi$ is positive when directed towards the centre of the rod, in which case $N_0$ is negative, i.e. compressive.
From the circular reference configuration, superimposed incremental deformations are analysed as induced by the application of an external incremental load $\dot{\mathbf{q}}^{\beta}$, to be detailed below. 
Assuming the inextensibility of the rod, the incremental kinematics is governed by 
\begin{equation}
\label{eq:dotdiffCirc}
    \frac{\partial^{5}{\dot{u}_{r}}}{\partial{\theta^{5}}} + \left(2 + \frac{{\Pi}R^{3}}{B}\right) \frac{\partial^{3}{\dot{u}_{r}}}{\partial{\theta^{3}}} + \left(1 + 2\frac{{\Pi}R^{3}}{B}\right) \frac{\partial{\dot{u}_{r}}}{\partial{\theta}} - \frac{\Pi R^3}{B} \dot{u}_\theta + \mathfrak{S} = 0 , 
    \quad
    \dot{u}_{r} + \frac{\partial \dot{u}_\theta}{\partial{\theta}} = 0 ,
\end{equation}
where a superimposed dot denotes an incremental quantity and the external load is specified through 
\begin{equation}
\label{eq:SgenCirc}
	\mathfrak{S} = -\frac{R^4}{B} \left(\frac{\partial\dot{\mathbf{q}}}{\partial\theta} \cdot {\mathbf{m}}_0 + 2 \dot{\mathbf{q}} \cdot \mathbf{t}_0\right) ,
\end{equation}
function of the incremental load $\dot{\mathbf{q}}$ applied to the rod.
The latter is not only due to $\dot{\mathbf{q}}^{\beta}$, but also contains components due to {\it both} the incremental interaction with the disk (which produces incremental traction when perturbed) {\it and} how the specific type of radial load $\Pi$ \lq reacts' to incremental deformation.
In particular, the annular rod enclosing the elastic disk is subject to a uniform radial load $\Pi$, selected between different types, which may be generated by the environment external to the disk/coating system or may be internally generated as a traction exchange between coating and disk, consequent to a shrink-fit or thermal operation. In particular, the radial force distribution $\Pi$ may be as follows.
\begin{itemize}
    \item Applied by the external environment on the outer surface of the coating. This $\Pi$ can be of three different natures: (i.) hydrostatic pressure, (ii.) centrally directed (towards the initial centre of the disk), and (iii.) dead.
    \item 
    Internally generated through traction exchange between coating and disk as induced by a preliminary shrink-fit process or differential variation of temperature between disk and coating. The account of this process enables the determination of the specific value of $\Pi$ to be used for incremental analysis where it is treated as a dead load, type (iii.) above.
\end{itemize}

The above loads define the reference configuration for the coated disk system, so that the rod is uniformly prestressed by an axial force, which can be either tensile or compressive, while the inner disk is either unloaded or slightly loaded through a uniform mean stress. This reference configuration is perturbed through the application of an additional incremental external load (preserving the overall equilibrium of the coating/disk system) $\dot{\mathbf{q}}^{\beta}$, with tangential and radial components $\dot{q}^{\beta}$ and $\dot{p}^{\beta}$ (Fig.~\ref{fig:Pload}), 
\begin{equation}
\label{eq:qdotSplit}
    \dot{\mathbf{q}}^{\beta} = \dot{q}^{\beta}\mathbf{t}_0 - \dot{p}^{\beta}\mathbf{m}_0 .
\end{equation}

The perturbation induces an incremental change in the reference configuration, so that the external load $\Pi$ and the internal tractions at the disk/coating contact also produce increments of loads for the rod. Summing up all these contributions, the incremental load for the rod, $\dot{\mathbf{q}}$, results as the sum 
\begin{equation}
\label{eq:qDotterm}
    \dot{\mathbf{q}} = \dot{\mathbf{q}}^{\Pi} + \dot{\mathbf{q}}^{\beta} + \dot{\mathbf{q}}^{\sigma} ,
\end{equation}
where, introducing $\mathscr{M}$ to rule the shear transmission properties at the interface ($\mathscr{M}=1$ for perfect bonding between disk and coating or $\mathscr{M}=0$ for slip contact), 
\begin{equation}
\label{eq:qdotSplit888}
    \dot{\mathbf{q}}^{\sigma} = -b\left(\dot{\sigma}_{rr}\mathbf{m}_0 + \mathscr{M}\dot{\sigma}_{r\theta}\mathbf{t}_0\right)_{r=R} ,
\end{equation}
is the incremental traction exchanged between disk and coating, where $b$ is the out-of-plane thickness of the coating and $\dot{\sigma}_{rr}$ and $\dot{\sigma}_{r\theta}$ are the incremental stress components on the boundary of the disk. The term $\dot{\mathbf{q}}^\Pi$ in eq.~\eqref{eq:qDotterm} describes the behaviour of the radial load during the increment, which may vary according to the specific dependence postulated for the force on the deformation. 
The following three incremental loadings are included in the formulation, each defining the term $\dot{\mathbf{q}}^\Pi$ as 
\begin{equation}
\label{eq:dorqPi}
    \dot{\mathbf{q}}^\Pi = 
    \Pi \times 
    \left\{
    \begin{array}{ll}
        \left(\dfrac{\partial{\dot{u}_{r}}}{\partial{s}} - \dfrac{\dot{u}_{\theta}}{R}\right)\mathbf{t}_{0} & \text{when $\Pi$ is a hydrostatic pressure (i.)} \\[4mm]
        -\dfrac{\dot{u}_\theta}{R}\,\mathbf{t}_0 & \text{when $\Pi$ is a centrally directed load (ii.)} \\[3mm]
        \bf{0} & \text{when $\Pi$ is a dead load (iii.)} 
    \end{array}
    \right.
\end{equation}
and hence, the term $\mathfrak{S}$ in equation \eqref{eq:SgenCirc} becomes 
\begin{equation}
\label{eq:SPi}
    \mathfrak{S}^{\Pi} = 
    \frac{\Pi R^3}{B} \times 
    \left\{
    \begin{array}{ll}
        -\dfrac{\partial\dot{u}_r}{\partial \theta} + \dot{u}_\theta & \text{when $\Pi$ is a hydrostatic pressure (i.)} \\[3mm]
        \dot{u}_\theta & \text{when $\Pi$ is a centrally directed load (ii.)} \\[3mm]
        0 & \text{when $\Pi$ is a dead load (iii.)} .
    \end{array}\right.
\end{equation} 

The hydrostatic pressure is a well-known type of load, which often can simply be realized, while a centrally directed load, passing through a fixed point (the centre of the circular ring, in the present case), requires a rather complicated realization, for instance, through inextensible cables. 
The most complicated loading condition for a circular geometry is the simpler for a rectilinear rod, namely, the dead loading. This can hardly be realized in a circular geometry, except when, instead of an external load, the prestress $N_0$ in the rod is generated through a thermal variation (in the presence of a mismatch in the thermal expansion coefficients of disk and rod) or a \lq shrink-fit' forcing of the coating on the disk. In this case, a residual stress, rather than a prestress, is generated, 
producing a radial pressure $\Pi$ on the coating, but the equations governing the problem become the same as the dead load, so that $\dot{\Pi}=0$.

\subsection{The disk coated with the prestressed rod}

\subsubsection{Incremental applied load on the disk/coating system}
\label{subsec:incrLoad}

Concerning the problem sketched in Fig.~\ref{fig:Pload}, the disk is characterized by a shear modulus $\mu^{\text{d}}$, Lam\'e constant $\lambda^{\text{d}}$, Young's modulus $E^{\text{d}}$, Poisson's ratio $\nu^{\text{d}}$ and it is coated along its boundary $L$ by the circular rod introduced in the previous section. 

In a polar coordinates system $(\mathbf{e}_{r},\mathbf{e}_{\theta})$, the incremental displacement of a point of the disk can be represented as
\begin{equation}
\label{eq:udisk}
    \dot{\mathbf{u}}^{\text{d}} = \dot{u}^{\text{d}}_{r}\, \mathbf{e}_r + \dot{u}^{\text{d}}_{\theta}\, \mathbf{e}_{\theta},
\end{equation}
where $u^{\text{d}}_r$ and $u^{\text{d}}_{\theta}$ are radial and tangential components.

Two initial stress configurations will be considered for the disk/coating system, before an additional incremental load, 
$\dot{\mathbf{q}}^\beta$ in Fig.~\ref{fig:Pload},
is applied on its boundary. These are:
\begin{enumerate}
    \item The radial load $\Pi$ is acting on the external surface of the coating and, as a consequence, a state of prestress is produced (in terms of the internal force $N_{0}$) in the coating, but inside of the disk the material remains unstressed, because the coating is modelled as an axially-inextensible rod, carrying $N_0$ without deformation.
    \item A uniform radial load is not applied on the external, but on the {\it internal} surface of the coating, where a radial force distribution $\Pi$ exists, acting on both coating and disk with an opposed sign, as a consequence of a previous \lq shrink/fit' or a thermal mismatch operation on the disk/coating system. This operation is assumed to generate a strong axial force $N_0$ in the coating, but to leave only a weak state of stress inside of the disk (that will be simply summed to further incremental stress).
\end{enumerate}

In the two above cases (1)-(2), the stress in the elastic disk is either null or assumed small before the incremental deformation occurs. In this circumstance, the equations of finite elasticity reduce to the linear theory, so that the incremental response of the disk is governed by Hooke's law 
\begin{equation}
\label{eq:dotSigma}
    \dot{\boldsymbol{\sigma}} = \lambda^{\text{d}}(\tr\mathbf{D})\,\mathbf{I} + 2\mu^{\text{d}}\, \mathbf{D}, 
\end{equation}
where $\mathbf{D} = (\nabla \dot{\mathbf{u}} - \nabla \dot{\mathbf{u}}^T)/2$ is the Eulerian strain incremental tensor, 
and the increments in the first Piola-Kirchhoff and Cauchy stresses coincide, $\dot{\mathbf{S}} = \dot{\boldsymbol{\sigma}}$.
Note that in the above case (2) the incremental solution has to be summed to the state of prestress in the disk, while in both cases the incremental solution is superimposed on the prestress in the annular coating.

In summary, the elastic disk is subject to incremental tractions, transmitted from its interaction with the coating. An incremental load, $\dot{\mathbf{q}}^\beta$ in Fig.~\ref{fig:Pload}, is applied to the latter. This load is summed to a previously-applied and uniformly-distributed radial load, $\Pi$ in Fig.~\ref{fig:Pload}, which may be induced by an either external or internal agent to the system. When applied from the external environment, the radial force distribution $\Pi$ does not produce any deformation in the disk and in the coating, due to the assumption of inextensibility of the latter, where an axial prestress, $N_0$ in Fig.~\ref{fig:Pload} is only generated. Alternatively, the state of prestress is produced by some shrink-fit operation, generating an internal radial load $\Pi$, which induces a prestress in the coating and is assumed to leave the elastic disk only weakly prestressed. 
In all cases, the effects of the prestress are implemented in the rod forming the coating, which obeys the linearized equations governing prestressed circular rods, while the elastic core inside the coating reacts linearly without a direct effect of prestress. The elastic core may be either fully connected to the coating or only radially bonded, to realize therefore a \lq slip contact'.

Implementing the incremental load (\ref{eq:qDotterm}) in equation \eqref{eq:dotdiffCirc} leads to the equations governing the incremental kinematics of the coating in contact with the disk
\begin{equation}
\label{eq:eqIncr}
    \begin{aligned}
        &\frac{\partial^{5}{\dot{u}^{\text{c}}_{r}}}{\partial{\theta^{5}}} + \left(2 + \frac{{\Pi}R^{3}}{B}\right)\frac{\partial^{3}{\dot{u}^{\text{c}}_{r}}}{\partial{\theta^{3}}} + \left(1 + 2\frac{{\Pi}R^{3}}{B}\right)\frac{\partial{\dot{u}^{\text{c}}_{r}}}{\partial{\theta}} - \frac{\Pi R^3}{B}\dot{u}^{\text{c}}_\theta + \mathfrak{S}^{\Pi} + \mathfrak{S}^{\sigma} + \mathfrak{S}^{\beta} = 0 , \\
        &\frac{\dot{u}^{\text{c}}_{r}}{R} + \frac{\partial \dot{u}^{\text{c}}_\theta}{\partial{s}} = 0 ,
    \end{aligned}
\end{equation}
where the superscript `c' stands for `coating' and
\begin{equation}
\label{eq:mPS}
    \mathfrak{S}^{\text{j}} = -\frac{R^4}{B}\left(\frac{\partial\dot{\mathbf{q}}^{\text{j}}}{\partial\theta} \cdot {\mathbf{m}}_0 + 2 \dot{\mathbf{q}}^{\text{j}} \cdot \mathbf{t}_0\right), \quad \text{j} = \Pi,\,\sigma,\,\beta .
\end{equation}
The derivatives involved in equation \eqref{eq:mPS} assume the forms
\begin{equation}
\label{eq:SsigmStar}
    \begin{aligned}
        &\frac{\partial\dot{\mathbf{q}}^{\sigma}}{\partial{s}} = -\left(\mathscr{M}\frac{\partial{\dot{\sigma}_{r\theta}}}{\partial{s}} + \frac{1}{R}\dot{\sigma}_{rr}\right)_{r=R} \mathbf{t}_0 + \left(\frac{\mathscr{M}}{R}\dot{\sigma}_{r\theta} - \frac{\partial{\dot{\sigma}_{rr}}}{\partial{s}}\right)_{r=R} \mathbf{m}_0 , \\[3mm] 
        &\frac{\partial\dot{\mathbf{q}}^{\beta}}{\partial{s}} = \left(\frac{\partial \dot{q}^{\beta}}{\partial s} - \frac{\dot{p}^{\beta}}{R}\right) \mathbf{t}_0 - \left(\frac{\partial \dot{p}^{\beta}}{\partial s} + \frac{\dot{q}^{\beta}}{R}\right) \mathbf{m}_0 ,
    \end{aligned}
\end{equation}
while the terms $\mathfrak{S}^{\sigma}$ and $\mathfrak{S}^{\beta}$ in equation \eqref{eq:mPS} become
\begin{equation}
\label{eq:msigmaStar}
    \mathfrak{S}^{\sigma} = \frac{R^{4}b}{B}\left( R\frac{\partial{\dot{\sigma}_{rr}}}{\partial{s}}+\mathscr{M}\dot{\sigma}_{r\theta}\right)_{r=R},
    \quad
    \mathfrak{S}^{\beta} = -\frac{R^4}{B}\left(-\frac{\partial \dot{p}^{\beta}}{\partial s} + \dot{q}^{\beta}\right) .
\end{equation}

When the coating is either perfectly bonded to the disk or radially connected but tangentially disconnected, the following boundary conditions (displacement continuity in the former case, partial continuity and vanishing of shear stress in the latter) have to be imposed, respectively, 
\begin{equation}
\label{eq:continuity}
    \dot{u}^{\text{c}}_{r} = \dot{u}^{\text{d}}_{r}\big{|}_{r=R} , 
    \quad
    \text{and}
    \quad
    \underbrace{\dot{u}^{\text{c}}_{\theta} = \dot{u}^{\text{d}}_{\theta}\big{|}_{r=R}}_{\text{perfect bonding}} 
    \quad 
    \text{or}
    \quad
    \underbrace{\dot{\sigma}_{r\theta}\big{|}_{r=R} = 0}_{\text{slip contact}} .
\end{equation}
Moreover,the governing equation \eqref{eq:eqIncr}$_1$ can be rewritten as
\begin{equation}
\label{eq:gov1all}
    \frac{\partial^{5}{\dot{u}^{\text{c}}_{r}}}{\partial{\theta^{5}}} + 2 \frac{\partial^{3}{\dot{u}^{\text{c}}_{r}}}{\partial{\theta^{3}}} + \frac{\partial{\dot{u}^{\text{c}}_{r}}}{\partial{\theta}} + \frac{\Pi R^3}{B}\left(\frac{\partial^{3}{\dot{u}^{\text{c}}_{r}}}{\partial{\theta^{3}}} + 2 \frac{\partial{\dot{u}^{\text{c}}_{r}}}{\partial{\theta}} - \dot{u}^{\text{c}}_\theta\right) + \mathfrak{S}^{\Pi} + \mathfrak{S}^{\sigma} + \mathfrak{S}^{\beta} = 0 ,
\end{equation}
which is valid for all load types listed in equation \eqref{eq:SPi} and for all the interface conditions specified with eqs.~\eqref{eq:continuity}, except for the combination of dead load and slip interface, when $\mathfrak{S}^{\Pi} = 0$ in equation \eqref{eq:eqIncr} and $\mathscr{M}=0$ in equation \eqref{eq:msigmaStar}$_1$. In the latter case, the following equation governs the problem
\begin{equation} 
\label{eq:deadSplip}
    \frac{\partial^{6}{\dot{u}^{\text{c}}_{r }}}{\partial{\theta^{6}}} + \left(2 + \frac{\Pi R^3}{B}\right)\frac{\partial^{4}{\dot{u}^{\text{c}}_{r}}}{\partial{\theta^{4}}} + \left(1+2\frac{\Pi R^3}{B}\right)\frac{\partial^{2}{\dot{u}^{\text{c}}_{r}}}{\partial{\theta^{2}}} + \frac{\Pi R^3}{B}\dot{u}^{\text{c}}_r + \frac{R^{5}}{B}\left(\frac{b}{R}\frac{\partial^{2}\dot{\sigma}_{rr}}{\partial{\theta^2}} + \frac{\partial^{2} \dot{p}^{\beta}}{\partial \theta^2} - \frac{\partial \dot{q}^{\beta}}{\partial \theta}\right) = 0 .
\end{equation}

\subsubsection{Complex variable formulation for the disk}
\label{subsec:cmplxDisk}

In a complex variable formulation, the disk is a simple connected circular region, bounded by a non-intersecting smooth curve $L$, so that every point can be represented through the complex variable $z = x_1 + i x_2$, where $x_1$ and $x_2$ are the coordinates of the point and $i = \sqrt{-1}$ is the imaginary unit. Moreover, denoting with $r$ the distance from the point $z$ to the origin $z_{c} = 0$ and with $\theta$ the angle (positive when anticlockwise) between $x_1$ and the radius $r$, in a polar coordinate system $(r, \theta)$ it is $z = re^{i\theta}$. Following \cite{mogilevskaya2018elastic}, the following notation is introduced
\begin{equation}
\label{eq:gz}
    \begin{aligned}
        &g(z) = \frac{R}{z} = \frac{R}{(x_{1} + i x_{2})} , 
        \quad 
        g^{\prime}(z) = -\frac{1}{R} g^{2}(z) , 
        \quad
        g^{\prime\prime}(z) = \frac{2}{R^{2}} g^{3}(z) , \\
        &\overline{g(z)} = \frac{R^{2}}{r^{2}} g^{-1}(z) , 
        \quad
        r = \sqrt{x_{1}^{2} + x_{2}^{2}}.
    \end{aligned}
\end{equation}
where a prime denotes the derivative with respect to $z$ and a superimposed bar the complex conjugate. By setting $r=R$ in equations \eqref{eq:gz}, the following relations for points $\tau = R\,e^{i\theta}$ on the boundary of the disk can be derived 
\begin{equation}
\label{eq:gtau}
    g(\tau) = \frac{R}{\tau} , 
    \quad 
    \overline{g(\tau)} = \frac{R}{\overline{\tau}} = g^{-1}(\tau) , 
    \quad
    g^{\prime}(\tau) = -\frac{1}{R} g^{2}(\tau) .
\end{equation}
The elastic displacement and stress fields can be determined everywhere in the disk via Kolosov-Muskhelishvili complex potentials $\varphi(z)$ and $\psi(z)$ as \cite{muskhelishvili2013some} 
\begin{equation}
\label{eq:muskhelishvilifields}
    \left\{
    \begin{aligned}
    & 2\mu^{\text{d}}{u^{\text{d}}(z)} = \kappa^{\text{d}}\varphi(z) - z\overline{\varphi^{\prime}(z)} - \overline{\psi(z)} , \\
	& \sigma^{\text{d}}_{11} + \sigma^{\text{d}}_{22} = 4\,\Rp\left(\varphi^{\prime}(z)\right) , \\
    & \sigma^{\text{d}}_{22} - \sigma^{\text{d}}_{11} + 2i\sigma^{\text{d}}_{12} = 2\left[\overline{z}\varphi^{\prime\prime}(z) + \psi^{\prime}(z)\right] ,
    \end{aligned} 
    \right.
\end{equation}
where the prime indicates derivation with respect to the variable $z$ while $\mathrm{Re}$ and $\mathrm{Im}$ denote real and the imaginary parts, respectively. The components of the incremental Eulerian strain tensor $\mathbf{D}$ are linked to the complex potentials through
\begin{equation}
\label{eq:muskhelishvilistrain}
    \begin{aligned}
        \left\{\begin{array}{lll}
		\displaystyle D^{\text{d}}_{11}+D^{\text{d}}_{22}=2 \frac{1-2\nu^{\text{d}}}{\mu^{\text{d}}} \mathrm{Re}\!\left(\varphi^{\prime}\left(z\right)\right), \\
		\displaystyle D^{\text{d}}_{22}-D^{\text{d}}_{11}+2iD^{\text{d}}_{12}=\frac{1}{\mu^{\text{d}}}\left[\overline{z}\varphi^{\prime\prime}\left(z\right)+\psi^{\prime}\left(z\right)\right]. 
		\end{array}
		\right.
	\end{aligned}
\end{equation}

At every point $\tau = R\, e^{i\theta}$ on the boundary of the disk the following complex Fourier representation for the displacement is introduced
\begin{equation}
\label{eq:cmplxFourierU}
    u^{\text{d}}_{1}(\tau) + i\, u^{\text{d}}_{2}(\tau) = \sum_{n=1}^{\infty}{A_{-n}\, g^{n}(\tau)} + \sum_{n=0}^{\infty}{A_{n}\, g^{-n}(\tau)} ,
\end{equation}

where $u^{\text{d}}_{1}(\tau)$ and $u^{\text{d}}_{2}(\tau)$ are displacement components and $A_{\pm{n}}$ are complex coefficients for the moment  unknown.
The radial and tangential components of the displacement are represented as
\begin{equation}
\label{eq:ur}
    u^{\text{d}}_{r}(\tau) = \frac{1}{2}\left[u^{\text{d}}\left(\tau\right)\, g\left(\tau\right) + \overline{u^{\text{d}}(\tau)}\, g^{-1}(\tau)\right] , 
    \quad 
    u^{\text{d}}_{\theta}(\tau) = \frac{1}{2i} \left[u^{\text{d}}(\tau)\, g(\tau) - \overline{u^{\text{d}}(\tau)}\, g^{-1}(\tau)\right] . 
\end{equation}
Eqns.~\eqref{eq:cmplxFourierU} and \eqref{eq:ur} lead to  
\begin{equation}
\label{urComp}
    \left.
    \begin{aligned}
        & 2u^{\text{d}}_{r}(\tau) \\
        & 2i\, u^{\text{d}}_{\theta}(\tau)
    \end{aligned}
    \right\} = 
    \sum_{n=1}^{\infty}{A_{-n}\, g^{n+1}(\tau)} + \sum_{n=0}^{\infty}{A_{n}\, g^{-\left(n-1\right)}(\tau)}\pm \sum_{n=1}^{\infty}{\overline{A_{-n}}\, g^{-\left(n+1\right)}(\tau)} \pm \sum_{n=0}^{\infty}{\overline{A_{n}}\, g^{n-1}(\tau)} .
\end{equation} 

The complex combination of the stresses acting at a point $\tau\in{L}$ of the disk can be introduced 
\begin{equation} 
\label{eq:traction}
    \sigma^{\text{d}}_{rr}(\tau) + i\, \sigma^{\text{d}}_{r\theta}(\tau) = \sum_{n=1}^{\infty}{B_{-n}\, g^{n}(\tau)} + \sum_{n=0}^{\infty}{B_{n}\, g^{-n}(\tau)} ,
\end{equation}
where the complex coefficients $A_{\pm{n}}$ and $B_{\pm{n}}$ are interrelated as \cite{zemlyanova2018circular} 
\begin{equation}
\label{ABinterrelation}
    \begin{aligned}
        & B_{-1} = 0 , 
        && B_{0} = \frac{4\mu^{\text{d}}}{\left(\kappa^{\text{d}}-1\right)R}\, \Rp\left(A_{1}\right) , \\
        & B_{-n} = \frac{2\mu^{\text{d}}}{R}(n-1)\, A_{1-n} , \text{ for } n \ge {2} , 
        && B_{n} = \frac{2\mu^{\text{d}}}{\kappa^{\text{d}}R}(n+1)\, A_{n+1} , \text{ for } n \ge {1} .
    \end{aligned}
\end{equation}
The applied external load is represented by the following complex series
\begin{equation}
\label{eq:Pcmplx}
    q^{\beta}(\tau) + i\,p^{\beta}(\tau) = \sum_{n=1}^{\infty}{D_{-n}\, g^{n}(\tau)} + \sum_{n=0}^{\infty}{D_{n}\, g^{-n}(\tau)} ,
\end{equation}
where $D_{\pm{n}}$ are complex coefficients that are known once the shape of the external load $\mathbf{q}^{\beta}$ is prescribed.

For a circular elastic disk, the expressions for the complex potentials $\varphi(z)$ and $\psi(z)$ appearing in eqs.~\eqref{eq:muskhelishvilifields} assume the form \cite{mogilevskaya2008multiple}
\begin{equation}
\label{eq:potentialdisc}
    \begin{aligned}
        & \varphi(z) = \frac{2\mu^{\text{d}}}{\kappa^{\text{d}}-1}\, \Rp\left(A_{1}\right)\, g^{-1}(z) + \frac{2\mu^{\text{d}}}{\kappa^{d}} \sum_{n=1}^{\infty}{A_{n+1}\, g^{-\left(n+1\right)}(z)} , \\
        & \psi(z) = -\frac{2\mu^{\text{d}}}{\kappa^{\text{d}}-1}\, \Rp\left(A_{1}\right)\, \frac{\overline{z_{c}}}{R} - \frac{2\mu^{\text{d}}}{\kappa^{\text{d}}} \left[\frac{\overline{z_{c}}}{R} + g(z)\right] \sum_{n=1}^{\infty}{\left(n+1\right)A_{n+1}\, g^{-n}(z)} \\
        & \phantom{\psi(z) = }-2\mu^{\text{d}} \sum_{n=2}^{\infty}{\overline{A_{1-n}}\, g^{-\left(n-1\right)}(z)} ,
    \end{aligned}
\end{equation}
and hence, the elastic fields on the boundary and within the disk are known once coefficients $A_{\pm{n}}$ are found as functions of the known coefficients $D_{\pm{n}}$. For brevity, only the case of \textit{perfect bonding condition} at the interface will be presented in the following, where conditions \eqref{eq:continuity}$_{1-2}$ are enforced. The other case of \textit{slip interface} will not be reported, but its derivation follows a procedure analogous to that developed for perfect bonding.

\section{Analytic solution for the disk with prestressed coating}
\label{sec:anSolution}

\subsection{Complex Fourier series form of the governing equations}
\label{subsec:cmplxEq}

The complex counterpart of equation \eqref{eq:dotdiffCirc}$_2$, representing the inextensibility constraint, can be determined using the series representation for the displacement, eq.~\eqref{urComp}, from which, collecting the terms with the same power of $g^{\pm{n}}(\tau)$, it follows \cite{gaibotti2022isoDisk}
\begin{equation}
\label{eq:inexsibility_cmplx}
    \Rp\left(A_{1}\right) = 0, 
    \quad 
    A_{2} = 0, 
    \quad 
    A_{n+1} = \frac{n-1}{n+1}\, \overline{A_{1-n}} 
    \quad 
    \text{for $n\neq{0}$ and $n\neq{-1}$} . 
\end{equation}

From equations \eqref{eq:SPi} and \eqref{eq:msigmaStar}$_1$ for $r=R$, adopting the Fourier series representation introduced before, the terms $\mathfrak{S}^{\Pi}$ and $\mathfrak{S}^{\sigma}$ in equation \eqref{eq:eqIncr}$_1$ become \cite{gaibotti2024bifurcations}
\begin{equation}
\label{eq:SpiSs_cmplx}
    \begin{aligned}
        \mathfrak{S}^{\Pi}(\tau) &= \xi\frac{\Pi R^{3}}{2i\,B}\left\{\sum_{n=1}^{\infty}(-n)^{\alpha}\left[A_{-n}\, g^{n+1}(\tau) - \overline{A_{-n}}\, g^{-\left(n+1\right)}(\tau)\right]\right. \\
        &\left. + \sum_{n=0}^{\infty}n^{\alpha}\left[A_{n}\, g^{-\left(n-1\right)}(\tau) - \overline{A_{n}}\, g^{n-1}(\tau)\right] - \left(\alpha-\xi\right) \left[A_0\, g(\tau) - \overline{A_0}\, g^{-1}(\tau)\right]\right\} , \\
        \mathfrak{S}^{\sigma}(\tau) &= \frac{R^{4}b}{2i\,B}\left\{\sum_{n=1}^{\infty}{\left(n+\mathscr{M}\right)}\left[B_{-n}\, g^{n}(\tau) - \overline{B_{-n}}\, g^{-n}(\tau)\right] - \sum_{n=0}^{\infty}{\left(n-\mathscr{M}\right)} \left[B_{n}\, g^{-n}(\tau) - \overline{B_{n}}\, g^{n}(\tau)\right]
        \right\} ,
    \end{aligned}
\end{equation}
where (i.) $\xi=\alpha=1$ for hydrostatic pressure, (ii.) $\xi=1$, $\alpha=0$ for centrally directed load, and (iii.) $\xi=0$ for dead load. Isolating the real and the imaginary parts in equation \eqref{eq:Pcmplx} yields
\begin{equation}
\label{eq:dotpq}
    \left.
    \begin{aligned}
        & 2\, \dot{p}^{\beta}(\tau) \\
        & 2i\, \dot{q}^{\beta}(\tau)
    \end{aligned}
    \right\} =
    \sum_{n=1}^{\infty}D_{-n}\, g^{n}(\tau) + \sum_{n=0}^{\infty}D_{n}\, g^{-n}(\tau) \pm \sum_{n=1}^{\infty}\overline{D_{-n}}\, g^{-n}(\tau) \pm \sum_{n=0}^{\infty}\, \overline{D_{n}}\, g^{n}(\tau),
\end{equation}
so that expression (96)$_2$, $d\tau/ds = i\, g^{-1}(\tau)$ in \cite{zemlyanova2018circular}, provides for the term $\mathfrak{S}^{\beta}$ in eq.~\eqref{eq:eqIncr}$_1$ 
\begin{equation}
\label{eq:Sstar_cmplx}
    \begin{aligned}
        \mathfrak{S}^{\beta}(\tau) &= \frac{R^4}{B}\left(\sum_{n=1}^{\infty}{(n-1)D_{-n}\, g^{n}(\tau)} - \sum_{n=0}^{\infty}(n+1)D_{n}\, g^{-n}(\tau)\right. \\
        &\phantom{=} \left. - \sum_{n=1}^{\infty}(n-1)\overline{D_{-n}}\, g^{-n}(\tau) + \sum_{n=0}^{\infty}{(n+1)\overline{D_{n}}\, g^{n}(\tau)}\right) .
    \end{aligned}
\end{equation}

From \cite{gaibotti2024bifurcations} the following terms appearing in eq.~\eqref{eq:gov1all} are identified as
\begin{equation}
\label{eq:cmplxTerms}
    \begin{aligned}
        \frac{\partial^5 \dot{u}_r}{\partial\theta^5} + 2 \frac{\partial^3 \dot{u}_r}{\partial\theta^3} + \frac{\partial \dot{u}_r}{\partial\theta} &= \frac{1}{2i} \left\{\sum_{n=1}^{\infty}n^{2}\left(n+1\right)\left(n+2\right)^{2} \left[A_{-n}\, g^{n+1} - \overline{A_{-n}}\, g^{-\left(n+1\right)}\right]\right. \\
        &\phantom{=} \left. - \sum_{n=3}^{\infty}{n^2\left(n-1\right)\left(n-2\right)^{2} \left[A_{n}\, g^{-\left(n-1\right)} - \overline{A_{n}}\, g^{n-1}\right]}\right\} , \\
        \frac{\partial^{3}{\dot{u}_{r}}}{\partial{\theta^{3}}} + 2 \frac{\partial{\dot{u}_{r}}}{\partial{\theta}} - \dot{u}_\theta &= \frac{1}{2i} \left\{\sum_{n=1}^{\infty} n\left(n^2-3n+1\right) \left[A_{n}\, g^{-\left(n-1\right)}(\tau) - \overline{A_{n}}\, g^{n-1}\right]\right. \\
        &\phantom{=} \left. - \sum_{n=1}^{\infty} n\left(n^2+3n+1\right) \left[A_{-n}\, g^{n+1} - \overline{A_{-n}}\, g^{-\left(n+1\right)}\right]\right\} ,
    \end{aligned}
\end{equation}
so that a substitution of expressions \eqref{eq:cmplxTerms}, \eqref{eq:SpiSs_cmplx} and \eqref{eq:Sstar_cmplx} leads to the following form of equation \eqref{eq:gov1all}
\begin{multline}
\label{eq:gov1all_cmplx}
    \sum_{n=1}^{\infty} n^{2}\left(n+1\right)\left(n+2\right)^{2} \left[A_{-n}\, g^{n+1}(\tau) - \overline{A_{-n}}\, g^{-\left(n+1\right)}(\tau)\right] \\
    - \sum_{n=3}^{\infty} {n^2\left(n-1\right)\left(n-2\right)^{2} \left[A_{n}\, g^{-\left(n-1\right)}(\tau) - \overline{A_{n}}\, g^{n-1}(\tau)\right]} + \frac{\Pi{R^3}}{B} \left\{\sum_{n=1}^{\infty} n\left(n^2-3n+1\right)\right. \\
    \left. \left[A_{n}\, g^{-\left(n-1\right)}(\tau) - \overline{A_{n}}\, g^{n-1}(\tau)\right] - \sum_{n=1}^{\infty} {n\left(n^2+3n+1\right) \left[A_{-n}\, g^{n+1}(\tau) - \overline{A_{-n}}\, g^{-\left(n+1\right)}(\tau)\right]}\right\} \\
    + \frac{bR^4}{B} \left\{\sum_{n=1}^{\infty} {\left(n+\mathscr{M}\right)} \left[B_{-n}\, g^{n}(\tau) - \overline{B_{-n}}\, g^{-n}(\tau)\right] - \sum_{n=0}^{\infty} {\left(n-\mathscr{M}\right)} \left[B_{n}\, g^{-n}(\tau) - \overline{B_{n}}\, g^{n}(\tau)\right]\right\} \\
    + \frac{R^4}{B} \left[\sum_{n=1}^{\infty} (n-1)D_{-n}\, g^{n}(\tau) - \sum_{n=0}^{\infty} {(n+1)D_{n}\, g^{-n}(\tau)} - \sum_{n=1}^{\infty} {(n-1)\overline{D_{-n}}\, g^{-n}(\tau)}\right. \\
    \left. + \sum_{n=0}^{\infty} {(n+1)\overline{D_{n}}}\, g^{n}(\tau)\right] + \mathfrak{S}^{\Pi} = 0 .
\end{multline}
Using eqs.~\eqref{ABinterrelation} and collecting terms with the same power in $g^{\pm{n}}(\tau)$ in equation \eqref{eq:gov1all_cmplx}, the following cases can be distinguished: 
\begin{itemize}
    \item For $n=0$ and $n=1$ 
        \begin{equation}
        \label{eq:Ima1}
            (\xi-1) \Pi\, \Ip(A_1) - R\, \Ip(D_0) = 0 ,
            \quad
            \Pi\, \xi(\alpha - \xi) A_0 + 2R\, \overline{D_1} = 0 ,
        \end{equation}
    \item For $n\geq{2}$
        \begin{equation}
        \label{eq:A1n_coeff}
            A_{1-n} = -\frac{R^4\kappa^{\text{d}}\left[(n-1)\, D_{-n} + (n+1)\, \overline{D_{n}}\right]}{2(n-1) \left[B \kappa^{\text{d}} n^2(n^2-1) - \Pi R^{3} \kappa^{\text{d}} \left(n^2 - 1 - \Upsilon(n,\alpha,\xi)\right) + b \mu^{\text{d}} R^{3} \Psi(n,\mathscr{M})\right]} ,
        \end{equation}
        where
        \begin{equation}
        \label{eq:upsVar}
            \Upsilon(n,\alpha,\xi) = \frac{\xi}{2} \left[(-1)^{\alpha}(n-1)^{\alpha-1} - (n+1)^{\alpha-1}\right] ,
            \quad
            \Psi(n,\mathscr{M}) = (n+\mathscr{M}) \kappa^{\text{d}} + n - \mathscr{M} .
        \end{equation}
\end{itemize}
The denominator in eq.~\eqref{eq:A1n_coeff}, for a radial compressive load, $\Pi>0$, may vanish and the corresponding elastic fields become singular. 
In particular, for a given set of material and geometric parameters $(E^{\text{c}},E^{\text{d}},\kappa^{\text{d}},R,b,J)$ a limit value $\Pi_{\text{cr}}$ exists for which the incremental solution for the disk/coating system bifurcates. 
After this limit value is exceeded, any solution for the coated disk is unstable and thus not anymore valid. 
The value of the dimensionless bifurcation radial load, as a function of the wave number $n$, was found in \cite{gaibotti2024bifurcations} as 
\begin{equation}
\label{eq:Picr}
    \frac{\Pi(n)R^3}{B} = \frac{n^{2}\left(n^2-1\right) + 
    \displaystyle  
    \frac{\mu^{\text{d}} b R^3}{\kappa^{\text{d}} B} \Psi(n,\mathscr{M})}{\left(n^2-1\right) - 
    \displaystyle 
    \Upsilon(n,\alpha,\xi)} ,
    \quad
    n \geq 2, 
\end{equation}
from which the critical value for the radial load $\Pi_{\text{cr}}$ corresponds to the integer number $n$ minimizing $\Pi (n)$.

\subsection{Elastic fields within the prestressed coated disk and internal forces in the coating}

Using eqs.~\eqref{eq:inexsibility_cmplx} and \eqref{eq:A1n_coeff} for the coefficients $A_{\pm{n}}$, the complex potentials and their derivatives involved in equations \eqref{eq:potentialdisc} assume the form
\begin{equation}
\label{eq:potentials}
    \begin{aligned}
        &\varphi(z) = \mu^{\text{d}}R^4 \sum_{n=2}^{\infty} \frac{1}{\Gamma(n+1)} \left[(n-1)\overline{D_{-n}} + (n+1)D_{n}\right]\, g^{-(n+1)}(z) , \\
        &\varphi^{\prime}(z) = \mu^{\text{d}}R^3 \sum_{n=2}^{\infty} \frac{1}{\Gamma} \left[(n-1)\overline{D_{-n}} + (n+1)D_{n}\right]\, g^{-n}(z) , \\
        &\varphi^{\prime\prime}(z) = \mu^{\text{d}}R^2 \sum_{n=2}^{\infty} \frac{1}{\Gamma}\, n\left[(n-1)\overline{D_{-n}} + (n+1)D_{n}\right]\, g^{-(n-1)}(z) , \\
        &\psi(z) = -\mu^{\text{d}}R^4 \sum_{n=2}^{\infty} \frac{1}{\Gamma(n-1)} \left[(n-1)\overline{D_{-n}} + (n+1)D_{n}\right](n-1+\kappa^{\text{d}})\, g^{-(n-1)}(z) , \\
        &\psi^{\prime}(z) = -\mu^{\text{d}}R^3 \sum_{n=2}^{\infty} \frac{1}{\Gamma} \left[(n-1)\overline{D_{-n}} + (n+1)D_{n}\right](n-1+\kappa^{\text{d}})\, g^{-(n-2)}(z) ,
    \end{aligned}
\end{equation}
where
\begin{equation}
\label{eq:denPhiPsi}
    \Gamma = B\kappa^{\text{d}} n^2(n^2-1) - \Pi R^3\kappa^{\text{d}} (n^2-1-\Upsilon(n,\alpha,\xi)) + b\mu^{\text{d}}R^3 \Psi(n,\mathscr{M}) .
\end{equation}

The elastic fields at every point $z$ within the disk can be evaluated by substituting eqs.~\eqref{eq:potentials} into the Kolosov-Muskhelishvili formulae \eqref{eq:muskhelishvilifields}. This leads to
\begin{equation}
\label{eq:muskComp}
    \begin{aligned}
        &u^{\text{d}}(z) = \frac{R^4}{2} \sum_{n=2}^{\infty} \frac{1}{\Gamma} \left\{-\frac{\kappa^{\text{d}} \left[(n-1)\overline{D_{-n}}+(n+1)D_{n}\right]\, g^{-(n+1)}(z)}{(n+1)}\right. \\
        & \qquad \left. + \frac{\left[(n-1)D_{-n}+(n+1)\overline{D_{n}}\right] \left[r^2(n-1) - R^2(n-1 + \kappa^{\text{d}})\right]R^{-2n}\, g^{n-1}(z)}{(n-1)r^{-(2n-2)}}\right\}, \\
        &\frac{\sigma^{\text{d}}_{11}(z)+\sigma^{\text{d}}_{22}(z)}{4\mu R^3} = \Rp\left[\frac{1}{\Gamma} \left[(n-1)\overline{D_{-n}} + (n+1)D_{n}\right]\, g^{-n}(z)\right] , \\
        &\frac{\sigma^{\text{d}}_{11}(z) - \sigma^{\text{d}}_{22}(z) + i\sigma^{\text{d}}_{12}(z)}{2\mu R^3} = -\sum_{n=2}^{\infty} \frac{1}{\Gamma}\left[(n-1)\overline{D_{-n}} + (n+1)D_{n}\right]\left[r^2n-R^2(n-1+\kappa^{\text{d}})\right]\, g^{-(n-2)}(z) .
\end{aligned}
\end{equation}

Internal forces in the prestressed coating loaded by the incremental applied load can be evaluated using equations (3.14) derived and reported in reference \cite{gaibotti2024bifurcations}, which are 
\begin{equation}
\label{eq:internalF}
    \begin{aligned}
        &\dot{M} = -B\left(\frac{\partial^{2}\dot{u}_r}{\partial s^2} + \frac{\dot{u}_r}{R^2}\right) , \\
        &\dot{T} = -B\left(\frac{1}{R^2}\frac{\partial\dot{u}_r}{\partial s} + \frac{\partial^{3}\dot{u}_r}{\partial s^3}\right) - \Pi R\left(\frac{\partial\dot{u}_r}{\partial s} - \frac{\dot{u}_{\theta}}{R}\right) , \\
        &\dot{N} = -R\left[B\left(\frac{\partial^{4}\dot{u}_r}{\partial s^4} + \frac{1}{R^2}\frac{\partial^2\dot{u}_r}{\partial s^{2}}\right) + \Pi R\left(\frac{\partial^{2}\dot{u}_r}{\partial s^2} + \frac{\dot{u}_r}{R^2}\right) + \dot{p}^{\beta} + b\dot{\sigma}_{rr}\right] .
    \end{aligned}
\end{equation}

\section{A case study: the coated disk subject to two opposite force distributions}
\label{subsec:case}

\subsection{Model for the applied external load}

Following \cite{gaibotti2022isoDisk}, the theoretical framework developed in the preview sections is now particularized to the case of a coated disk loaded by two opposite self-equilibrated force distributions $\dot{p}^\beta$, of incremental nature, applied along an arc length $s = \gamma R$, where $\gamma$ is the angle centred along the upper and lower part of the vertical diameter of the disk (see the inset in Fig.~\ref{fig:uP}). 
From expression \eqref{eq:qDotterm}$_1$, assuming $\dot{q}^{\beta}=0$, the value of the complex coefficients $D_{\pm{n}}$ involved in the complex Fourier representation of the applied incremental load, equation \eqref{eq:Pcmplx}, can be generated using the equation (68) derived and reported in \cite{gaibotti2022isoDisk}, which now becomes
\begin{equation}
\label{eq:fourierD}
    \int_{\frac{\pi-\gamma}{2}}^{\frac{\pi+\gamma}{2}} \dot{p}^{\beta} e^{-mi\theta}\, d\theta + \int_{\frac{3\pi-\gamma}{2}}^{\frac{3\pi+\gamma}{2}} \dot{p}^{\beta} e^{-mi\theta}\, d\theta = \int_{0}^{2\pi} \left[\sum_{n=1}^{\infty} D_{-n}\,e^{-(n+m)\,i\theta} + \sum_{n=0}^{\infty} D_{n}\, e^{-(n-m)\, i\theta}\right]\, d\theta .
\end{equation}
For a fixed integer $m$ in equation \eqref{eq:fourierD} one non-vanishing coefficient is generated and its value can be computed by inverting the same equation, which gives
\begin{equation}
\label{eq:fourierD2}
    D_m = \frac{1}{2\pi} 
    \left\{ 
    \int_{\frac{\pi-\gamma}{2}}^{\frac{\pi+\gamma}{2}} \dot{p}^{\beta} e^{-mi\theta}\, d\theta + \int_{\frac{3\pi-\gamma}{2}}^{\frac{3\pi+\gamma}{2}} \dot{p}^{\beta} e^{-mi\theta}\, d\theta
    \right\}.
\end{equation}
In the case that the applied incremental loading $\dot{p}^\beta$ is constant, the above equation reduces to the explicit form 
\begin{equation}
\label{eq:fourierD3}
    \begin{aligned}
        &D_m = i \frac{\dot{p}^\beta}{2m\pi} \left(1 + e^{i m \pi} \right) \left(1 - e^{i m \gamma} \right) e^{-i\frac{m}{2}(3\pi + \gamma)}, && m \neq 0, \\[3mm]
        &D_0 = \dot{p}^{\beta}\frac{\gamma}{\pi}.
    \end{aligned}
\end{equation}

\subsection{Fixing the rigid body roto-translations}

Once the coefficients $D_{\pm{n}}$ are generated, the value of the complex coefficients $A_{\pm{n}}$ can be derived from eq.~\eqref{eq:A1n_coeff}. However, the coefficient $A_0$ and the imaginary part of $A_1$ remain unknown as they rule rigid body roto-translations, so that their expressions can be computed, following the same procedure adopted and described in \cite{mogilevskaya2008multiple}.

When the prestress in the coating is generated by a radial loading of the \lq central direction' type ($\xi=1$, $\alpha=0$), a rigid rotation is a solution, while rigid translations are not. For dead radial load ($\xi=0$), rigid translations are solutions, but not rigid rotations, \cite{singer1970buckling}. Finally, for hydrostatic pressure ($\xi=\alpha=1$) rigid rotations and translations are always solutions. 
In these three cases of different applied radial loads, from equations \eqref{eq:Ima1}, requirements about the values of $A_0$ and $\mathrm{Im}(A_1)$ are found and reported in Table \ref{tab:rgbA}. 
\begin{table}[h!]
    \centering
	\begin{tabular}{@{} c@{\hspace{4\tabcolsep}} c@{\hspace{3\tabcolsep}} c@{\hspace{4\tabcolsep}} c@{} @{}}
    \toprule
    & Hydrostatic pressure & Centrally directed & Dead \\
    & $\xi=\alpha=1$ & $\xi=1,\ \alpha=0$ &$\xi=0$ \\
    \midrule
    $A_0$ & \mbox{unrestricted} & $-\frac{2R}{\Pi}\overline{D_1}$ & \mbox{unrestricted} \\[5pt]
    $\Ip(A_1)$ & \mbox{unrestricted} & \mbox{unrestricted} & $-R\,\mathrm{Im}(D_0)$ \\
    \bottomrule
	\end{tabular}
    \par
	\captionof{table}{\label{tab:rgbA}{Coefficients $A_0$ and $\Ip(A_1)$ determined to exclude rigid body roto-translations of the coated disk under applied load.}}
\end{table}

When coefficients $A_0$ and/or $\Ip(A_1)$ are unrestricted, their expressions are obtained by imposing displacements as derived and reported in \cite{gaibotti2022isoDisk}, so obtaining $A_0=0$ and the following expression for the imaginary part of $A_1$ 
\begin{equation}
\label{eq:ImA1rgb}
    \Ip(A_1) = -2\Ip\left(\frac{1}{n+1}A_{1-n}\right) .
\end{equation}

\subsection{Results for the coated disk at different levels of prestress, subject to opposite force distributions}

All the three different types of radial loads considered in the present article, equation (\ref{eq:dorqPi}), are investigated, to generate in the coating the same level of prestress (in the figures these are labelled as \lq Hydrostatic', \lq Centrally-directed', and \lq Dead'). The prestress is assumed to be a fraction of the critical load for bifurcation $\Pi_{\text{cr}}$, equation \eqref{eq:Picr}.  

In particular, four values of prestress are analysed: 
\begin{itemize}
    \item[(i.)] null, $\Pi=0$;
    \item[(ii.)] low, $\Pi/\Pi_{\text{cr}}=0.1$;
    \item[(iii.)] medium, $\Pi/\Pi_{\text{cr}}=0.5$;
    \item[(iv.)] close to the critical load of the disk/coating system, $\Pi/\Pi_{\text{cr}}=0.8$. 
\end{itemize}

Upon the application of any of the radial loads, the inner disk remains unloaded, so that the response of the coating/disk system is perturbed by applying an additional incremental load $\dot{p}^\beta$ uniformly distributed on a small arc, assumed of 4$^\circ$ and approximated using the first 100 terms in the series representation, eq.~\eqref{eq:fourierD3}. 

The incremental solution can be expressed in a dimensionless form as
\[
\frac{\mathbf{u}}{R} \quad \text{and} \quad \frac{\boldsymbol{\sigma}}{\mu^{\text{d}}},
\]
so that these quantities depend upon 3 nondimensional parameters
\[
\frac{B}{\mu^{\text{d}} b R^3}, \quad \frac{\Pi}{\mu^{\text{d}} b}, \quad \text{and} \quad \kappa^{\text{d}} .
\]

\begin{figure}[htb!]
    \centering
    \includegraphics[scale=0.97,keepaspectratio]{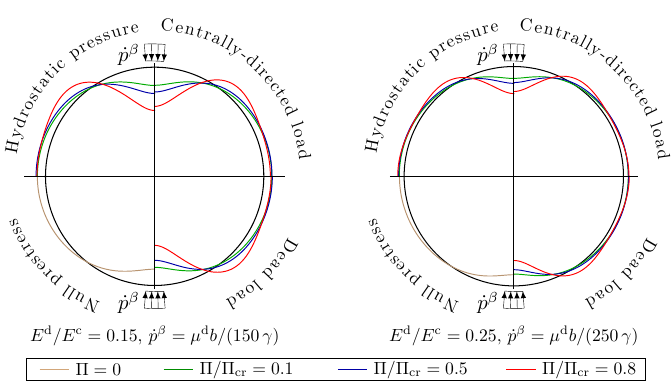}
    \caption{Incrementally deformed shape of the coating for two values of the ratio $E^{\text{d}}/E^{\text{c}} = \{0.15, 0.25\}$, upon application of the incremental load $\dot{p}^\beta = \{\mu^{\text{d}} b /(150 \gamma), \mu^{\text{d}} b /(250 \gamma)\}$, respectively, in the presence of a prestress of different intensity, $\Pi/\Pi_{\text{cr}}$ (0.1 green, 0.5 blue and 0.8 red line). Different types of radial loads are considered to produce the same prestress: hydrostatic pressure, centrally directed and dead load. The brown line refers to the case of the coated disk without prestress.}
    \label{figone}
\end{figure}

\begin{figure}[htb!]
    \centering
    \includegraphics[scale=0.97,keepaspectratio]{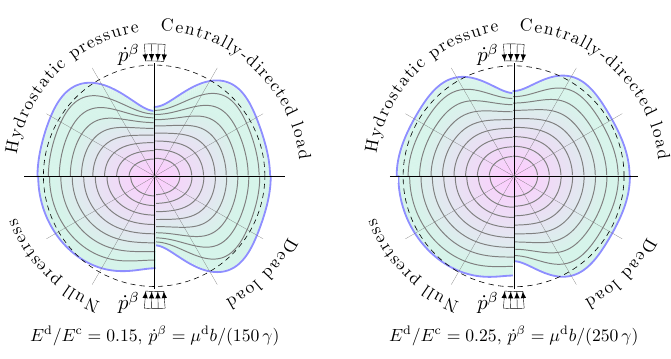}
    \caption{Incrementally deformed shape of the coating for two values of the ratio $E^{\text{d}}/E^{\text{c}} = \{0.15, 0.25\}$, upon application of the incremental load $\dot{p}^\beta = \{\mu^{\text{d}} b /(150 \gamma), \mu^{\text{d}} b /(250 \gamma)\}$, respectively, in the presence of a prestress $\Pi/\Pi_{\text{cr}}=0.8$. Different types of radial loads are considered to produce the same prestress: hydrostatic pressure, centrally directed and dead load. Level sets of the displacements are reported internally to the disk.}
    \label{fig:ut}
\end{figure}

The examples below are limited for brevity to the case of perfect bonding between coating and disk and the Kolosov constant is fixed as $\kappa^{\text{d}} = 2$. Moreover, rather than express the results as functions of the remaining two of the above parameters, it was decided to introduce the more intuitive ratio between Young's moduli of the disk and coating, $E^{\text{d}}/E^{\text{c}}$, and to assume $2(1+\nu^{\text{d}}) J/(bR^3) = 0.001$. The results are reported in Figs.~\ref{figone}--\ref{fig:ut}. 
%
%

The external deformed shape of the coated disk is reported in Figs.~\ref{figone} and \ref{fig:ut}, for different values of prestress in the former case, at fixed prestress $\Pi/\Pi_{\text{cr}}=0.8$ in the latter, where the deformed inner circles are also reported (the colour is proportional to the intensity of the field). In both cases two values of the ratio $E^{\text{d}}/E^{\text{c}}$ are considered, namely $E^{\text{d}}/E^{\text{c}}$ equal 0.15 and 0.25.

Both figures show that the compressive prestress decreases the stiffness of the system, which tends to vanish when the prestress in the coating approaches the bifurcation $\Pi/\Pi_{\text{cr}}=1$. 
The reported examples are all compared for the same fraction of prestress compared to the critical value, so that the differences are due to the particular type of radial load. The latter strongly affects the incremental deformation and the shape of the incrementally deformed solid. 

\begin{figure}[htb!]
    \centering
    \includegraphics[scale=0.97,keepaspectratio]{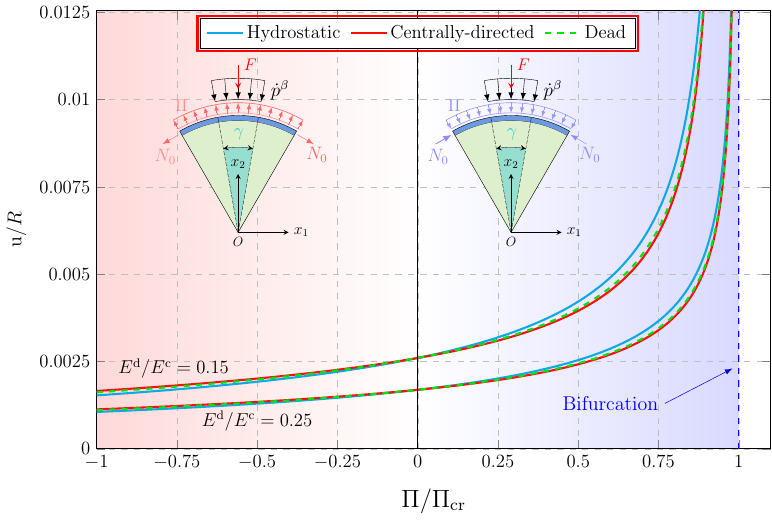}
    \caption{Incremental displacement (divided by $R$ and positive when towards the centre of the disk) of the point on the coating located under the resultant $F$ of the incremental load $\dot{p}^\beta$ (distributed on an arch $\gamma=4^\circ$) as a function of the prestress $\Pi/\Pi_{\text{cr}}$, ranging from tensile (negative values) to compressive (positive values). The differently coloured curves refer to different types of radial loads, all providing the same prestress ratio $\Pi/\Pi_{\text{cr}}$ in the coating. The asymptote of the curves denotes the critical load for bifurcation.}
    \label{fig:uP}
\end{figure}

The incremental displacement at the centre of the coating (under the resultant $F$ of the applied incremental load) 
is reported in Fig.~\ref{fig:uP} as a function of the prestress $\Pi/\Pi_{\text{cr}}$, including also the tensile case (negative values of $\Pi$). Results are given for $E^{\text{d}}/E^{\text{c}} = \{0.15, 0.25\}$. 
%

\begin{figure}[hbt!]
\centering
    \includegraphics[scale=0.97,keepaspectratio]{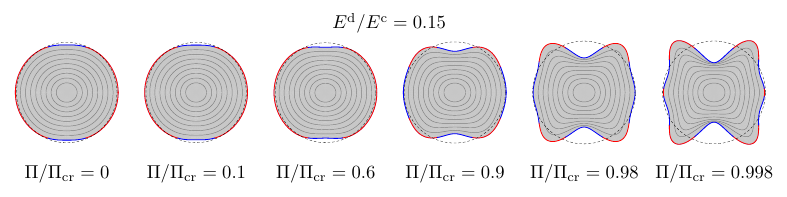}
    \includegraphics[scale=0.97,keepaspectratio]{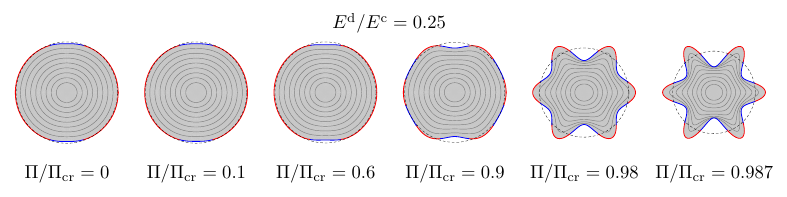}
    \caption{Incrementally deformed configurations of the coated disk for different ratios of $\Pi/\Pi_{\text{cr}}$ (from null prestress on the left to near-buckling pre-stress on the right) and for $E^{\text{d}}/E^{\text{c}}=0.15$ (upper part) and $E^{\text{d}}/E^{\text{c}}=0.25$ (lower part). The cases reported correspond to a pre-stress in the coating generated by hydrostatic pressure and the blue (red) colour highlights compressive (tensile) incremental tractions in the coating. Note that in the lower part, the bifurcation mode $n=6$ is approached, while in the upper part, the deformation remains symmetric even if the bifurcation mode is odd, $n=5$.} 
\label{tettone}
\end{figure}

The figure shows that the first bifurcation load is obtained in a perturbative way, so that the critical load corresponds to the asymptote of the graphs. In particular, the critical bifurcation occurs at $n=5$ for $E^{\text{d}}/E^{\text{c}} = 0.15$ and $n=6$ for $E^{\text{d}}/E^{\text{c}} = 0.25$. The situation is detailed in Fig.~\ref{tettone}, referred only to the case of hydrostatic pressure load. Here, at the increase of the prestress, the first bifurcation load is approached for both values of $E^{\text{d}}/E^{\text{c}}$, but for $E^{\text{d}}/E^{\text{c}} = 0.25$ the deformed mode approaches the bifurcation mode $n=6$, while in the other case the bifurcation mode is not correctly approached and the deformation remains symmetric, although $n=5$. However, in the latter case, when the bifurcation prestress is surpassed, the stiffness of the system becomes negative, thus showing that the critical load has been exceeded. 
%

\begin{figure}[hbt!]
\centering
    \includegraphics[keepaspectratio]{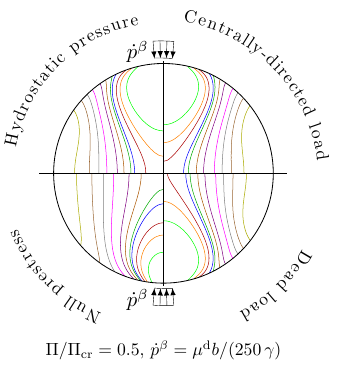}
    \includegraphics[keepaspectratio]{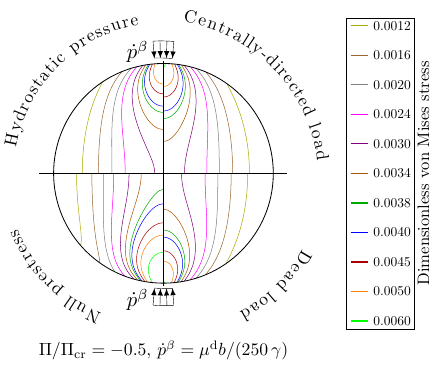}
    \caption{Level sets for dimensionless von Mises stress inside a coated disk, induced by the application of the incremental load $\dot{p}^\beta$, superimposed to a radial load producing the prestress in the coating. The prestress (compressive on the left and tensile on the right) in the coating is induced by different radial loadings. Results are compared to the case of the coated disk without prestressed for $E^{\text{d}}/E^{\text{c}}=0.25$. Note that the coating shields the stress inside the disk particularly when the prestress is tensile.} 
\label{tettine}
\end{figure}

The level sets of the dimensionless von Mises stress inside the disk generated by the application of the incremental load are depicted in Fig.~\ref{tettine}, for $E^{\text{d}}/E^{\text{c}} = 0.25$ and two values of prestress (compressive $\Pi/\Pi_{\text{cr}} = 0.5$ and tensile $\Pi/\Pi_{\text{cr}}=-0.5$).
%
%
In the figure identical colours correspond to the same level of von Mises stress in all quadrants and for both figures on the right and the left. Results highlight the fact that a coating subject to tensile prestress produces a shield to the inner core, which is less stressed than in the case in which the prestress is absent.

\section{Conclusions}

States of prestress or residual stress in solids can originate from different sources, such as thermal mismatch, shrink-fit operations, but also external pre-loads, namely, loads already acting before any further increment of load is superimposed. The mentioned sources play an important role in the incremental behaviour of a solid, as shown in the present article through the investigation of the response of an elastic disk with prestressed coating, the latter modelled through an elastic circular (unshearable and inextensible) rod. 
Based on the main assumptions that the prestress state is negligible in the disk but affects the coating and that the latter is axially inextensible, 
an incremental solution, found analytically via complex potentials, has been proposed when the prestressed coated disk is subjected to an arbitrary incremental load distribution. 
The presented results show the importance of several effects related to: (i.) the type of radial load acting to generate the prestress before the incremental load is applied; (ii.) the interfacial conditions between coating and bulk solid; (iii.) the stiffening or weakening induced by prestress; (iv.) the possibility of approaching bifurcation loads and modes through a perturbative technique.

The use of stiff coatings is common in several man-made and natural systems, where interfacial conditions, type of loading, and state of prestress play an important role. 
The presented results may find applications in biomechanics and in the field of deformable solids used for mechanical actuation or load bearing.

\clearpage

\section*{Acknowledgements}
The work has been developed in the framework of a NSF-ERC visit of S.M. (based in the United States) to the ERC project 101052956-Beyond, managed by D.B.. 
D.B. and M.G. gratefully acknowledge funding from the European Research Council (ERC) under the European Union’s Horizon 2020 research and innovation programme, Grant Agreement No. ERC-ADG-2021-101052956-BEYOND. 
A.P. gratefully acknowledges funding from the European Union (ERC CoG 2022, SFOAM, 101086644). 
S.M. gratefully acknowledges the support from the National Science Foundation, United States, award number NSF CMMI-2112894.

\printbibliography

@article{armenakas1963vibrations,
  title={Vibrations of infinitely long cylindrical shells under initial stress},
  author={Armenakas, Anthony E and Herrmann, G},
  journal={AIAA journal},
  volume={1},
  number={1},
  pages={100--106},
  year={1963}
}

@article{bigoni2008dynamics,
  title={Dynamics of a prestressed stiff layer on an elastic half space: filtering and band gap characteristics of periodic structural models derived from long-wave asymptotics},
  author={Bigoni, Davide and Gei, Massimiliano and Movchan, AB},
  journal={Journal of the Mechanics and Physics of Solids},
  volume={56},
  number={7},
  pages={2494--2520},
  year={2008},
  publisher={Elsevier}
}

@article{ogdencino,
  title={The effect of elastic surface coating on the finite deformation and bifurcation of a pressurized circular annulus},
  author={R.W. Ogden and D. Steigmann and D. Haughton},
  journal={Journal of elasticity},
  volume={47},
  number={},
  pages={121--145},
  year={1997},
  publisher={Elsevier}
}

@article{beuth1992cracking,
  title={Cracking of thin bonded films in residual tension},
  author={Beuth Jr, JL},
  journal={International Journal of Solids and Structures},
  volume={29},
  number={13},
  pages={1657--1675},
  year={1992},
  publisher={Elsevier}
}

@article{cai2000exact,
  title={Exact and asymptotic stability analyses of a coated elastic half-space},
  author={Cai, Zongxi and Fu, Yibin},
  journal={International journal of solids and structures},
  volume={37},
  number={22},
  pages={3101--3119},
  year={2000},
  publisher={Elsevier}
}

@article{chen2022review,
  title={Review on residual stresses in metal additive manufacturing: formation mechanisms, parameter dependencies, prediction and control approaches},
  author={Chen, Shu-guang and Gao, Han-jun and Wu, Qiong and Gao, Zi-han and Zhou, Xin and others},
  journal={Journal of Materials Research and Technology},
  volume={17},
  pages={2950--2974},
  year={2022},
  publisher={Elsevier}
}

@article{ogden84,
  title={Plane strain dynamics of elastic solids with intrinsic boundary elasticity, with application to surface wave propagation},
  author={Ray W Ogden and David J Steigmann},
  journal={Journal of the Mechanics and Physics of Solids},
  volume={50},
  pages={1869--1896},
  year={2002},
  publisher={Elsevier}
}

@article{gei,
  title={Elastic waves guided by a material interface},
  author={Massimiliano Gei},
  journal={European Journal of Mechanics},
  volume={27},
  pages={328--345},
  year={2008},
  publisher={Elsevier}
}

@article{gaibotti2022isoDisk,
	title = {Elastic disk with isoperimetric Cosserat coating},
	journal = {European Journal of Mechanics - A/Solids},
	pages = {104568},
	year = {2022},
	issn = {0997-7538},
	doi = {https://doi.org/10.1016/j.euromechsol.2022.104568},
	url = {https://www.sciencedirect.com/science/article/pii/S0997753822000559},
	author = {Matteo Gaibotti and Davide Bigoni and Sofia G. Mogilevskaya},
}

@article{gaibotti2024bifurcations,
  title={Bifurcations of an elastic disc coated with an elastic inextensible rod},
  author={Gaibotti, M and Mogilevskaya, SG and Piccolroaz, A and Bigoni, D},
  journal={Proceedings of the Royal Society A},
  volume={480},
  number={2281},
  pages={20230491},
  year={2024},
  publisher={The Royal Society}
}

@article{gei2002vibration,
  title={Vibration of a surface-coated elastic block subject to bending},
  author={Gei, Massimiliano and Ogden, RW},
  journal={Mathematics and Mechanics of Solids},
  volume={7},
  number={6},
  pages={607--628},
  year={2002},
  publisher={Sage Publications Sage CA: Thousand Oaks, CA}
}

@article{holzapfel2010constitutive,
  title={Constitutive modelling of arteries},
  author={Holzapfel, Gerhard A and Ogden, Ray W},
  journal={Proceedings of the Royal Society A: Mathematical, Physical and Engineering Sciences},
  volume={466},
  number={2118},
  pages={1551--1597},
  year={2010},
  publisher={The Royal Society Publishing}
}

@article{humphrey2003continuum,
  title={Continuum biomechanics of soft biological tissues},
  author={Humphrey, Jay D},
  journal={Proceedings of the Royal Society of London. Series A: Mathematical, Physical and Engineering Sciences},
  volume={459},
  number={2029},
  pages={3--46},
  year={2003},
  publisher={The Royal Society}
}

@article{hoger1985residual,
  title={On the residual stress possible in an elastic body with material symmetry},
  author={Hoger, Anne},
  journal={Archive for Rational Mechanics and Analysis},
  volume={88},
  pages={271--289},
  year={1985},
  publisher={Springer}
}

@article{hoger1986determination,
  title={On the determination of residual stress in an elastic body},
  author={Hoger, Anne},
  journal={Journal of Elasticity},
  volume={16},
  number={3},
  pages={303--324},
  year={1986},
  publisher={Springer}
}

@article{jensen1990decohesion,
  title={Decohesion of a cut prestressed film on a substrate},
  author={Jensen, Henrik M and Hutchinson, John W and Kyung-Suk, Kim},
  journal={International Journal of Solids and Structures},
  volume={26},
  number={9-10},
  pages={1099--1114},
  year={1990},
  publisher={Elsevier}
}

@article{jorgensen1995cracking,
  title={The cracking and spalling of multilayered chromium coatings},
  author={J{\o}rgensen, O and Horsewell, A and S{\o}rensen, Bent F and Leisner, Peter},
  journal={Acta metallurgica et materialia},
  volume={43},
  number={11},
  pages={3991--4000},
  year={1995},
  publisher={Elsevier}
}

@article{man1987towards,
  title={Towards an acoustoelastic theory for measurement of residual stress},
  author={Man, Chi-Sing and Lu, WY},
  journal={Journal of elasticity},
  volume={17},
  number={2},
  pages={159--182},
  year={1987},
  publisher={Springer}
}

@article{merodio2013influence,
  title={The influence of residual stress on finite deformation elastic response},
  author={Merodio, Jos\'{e} and Ogden, Ray W and Rodr\'{i}guez, Javier},
  journal={International Journal of Non-Linear Mechanics},
  volume={56},
  pages={43--49},
  year={2013},
  publisher={Elsevier}
}

@article{mogilevskaya2008multiple,
  title={Multiple interacting circular nano-inhomogeneities with surface/interface effects},
  author={Mogilevskaya, Sofia G and Crouch, Steven L and Stolarski, Henryk K},
  journal={Journal of the Mechanics and Physics of Solids},
  volume={56},
  number={6},
  pages={2298--2327},
  year={2008},
  publisher={Elsevier}
}

@article{mogilevskaya2018elastic,
	title={On the elastic far-field response of a two-dimensional coated circular inhomogeneity: Analysis and applications},
	author={Mogilevskaya, Sofia G and Zemlyanova, Anna Y and Zammarchi, Mattia},
	journal={International Journal of Solids and Structures},
	volume={130},
	pages={199--210},
	year={2018},
	publisher={Elsevier}
}

@book{muskhelishvili2013some,
	title={Some basic problems of the mathematical theory of elasticity},
	author={Muskhelishvili, N.I.},
	year={1959},
	publisher={Springer Science \& Business Media}
}

@article{noyan1991residual,
  title={Residual stresses in materials},
  author={Noyan, IC and Cohen, JB},
  journal={American Scientist},
  volume={79},
  number={2},
  pages={142--153},
  year={1991},
  publisher={JSTOR}
}

@article{shams2011initial,
  title={Initial stresses in elastic solids: constitutive laws and acoustoelasticity},
  author={Shams, Moniba and Destrade, Michel and Ogden, Ray W},
  journal={Wave Motion},
  volume={48},
  number={7},
  pages={552--567},
  year={2011},
  publisher={Elsevier}
}

@article{singer1970buckling,
  title={On the buckling of rings under constant directional and centrally directed pressure},
  author={Singer, J and Babcock, CD},
  journal={ Journal of Applied Mechanics},
  year={1970}
}

@article{zemlyanova2018circular,
	title={Circular inhomogeneity with Steigmann--Ogden interface: Local fields, neutrality, and Maxwell’s type approximation formula},
	author={Zemlyanova, A.Y.  and Mogilevskaya, S.G},
	journal={International Journal of Solids and Structures},
	volume={135},
	pages={85--98},
	year={2018},
	publisher={Elsevier}
}

\end{document}